\documentclass[prd,twocolumn,amssymb,showpacs,floatfix,%
superscriptaddress]{revtex4}

\usepackage{graphicx}

\def\fullfigwidth{9.3cm}
\def\smallfigwidth{7.3cm}
\def\ud{\,\mathrm{d}}

\begin{document}

\title{Theoretical Support for the Hydrodynamic Mechanism of Pulsar Kicks}

\author{J.~Nordhaus}
\email{nordhaus@astro.princeton.edu; tbrandt@astro.princeton.edu; burrows@astro.princeton.edu; livne@phys.huji.ac.il; cott@tapir.caltech.edu}
\affiliation{Department of Astrophysical Sciences, Princeton University, Princeton, NJ 08544, U.S.A.}

\author{T. D. Brandt}
\affiliation{Department of Astrophysical Sciences, Princeton University, Princeton, NJ 08544, U.S.A.}

\author{A. Burrows}
\affiliation{Department of Astrophysical Sciences, Princeton University, Princeton, NJ 08544, U.S.A.}

\author{E. Livne}
\affiliation{Racah Institute of Physics, Hebrew University, Jerusalem, Israel}

\author{C. D. Ott}
\affiliation{Theoretical Astrophysics, Mail Code 350-17, California Institute of Technology, Pasadena, CA 91125 U.S.A.}

\date{\today}

\begin{abstract}
The collapse of a massive star's core, followed by a neutrino-driven, asymmetric supernova explosion, can naturally lead to pulsar recoils and neutron star kicks.  Here, we present a two-dimensional, radiation-hydrodynamic simulation in which core collapse leads to significant acceleration of a fully-formed, nascent neutron star (NS) via an induced, neutrino-driven explosion.  During the explosion, a $\sim$10\% anisotropy in the low-mass, high-velocity ejecta lead to recoil of the high-mass neutron star.  At the end of our simulation, the NS has achieved a velocity of $\sim$150 km$\,$s$^{-1}$ and is accelerating at $\sim$350 km$\,$s$^{-2}$, but has yet to reach the ballistic regime.  The recoil is due almost entirely to hydrodynamical processes, with anisotropic neutrino emission contributing less than 2\% to the overall kick magnitude.  Since the observed distribution of neutron star kick velocities peaks at $\sim$300-400 km$\,$s$^{-1}$, recoil due to anisotropic core-collapse supernovae provides a natural, non-exotic mechanism with which to obtain neutron star kicks.
\end{abstract}


\maketitle

\section{\label{sec:intro}Introduction\protect}

The velocity distribution of young pulsars bears little resemblance to
that of their massive star progenitors \cite{Gunn:1970ys}.  Typical
birth velocities range from $\sim$200-500 km$\,$s$^{-1}$, with some
reaching upwards of $\sim$1000 km$\,$s$^{-1}$ \cite{Chatterjee:2005uq}.
While the observed pulsar velocities may hint at a two-component
distribution (possibly implying two populations)
\cite{Cordes:1998lq,Brisken:2003kx,Arzoumanian:2002fj}, recent work
supports a single, Maxwellian distribution
\cite{Lyne:1994rr,Hansen:1997dq,Hobbs:2005fk,Zou:2005yq,Faucher-Giguere:2006lr}.

Various mechanisms for the origin of neutron star kicks and pulsar
recoil and their connections with pulsar spins have been proposed
\cite{Spruit:1998fj}.  Misaligned jet/counter-jets during the
supernova explosion might produce sufficient acceleration if they
are launched near the proto-neutron star (PNS)
\cite{Cen:1998oq,Khokhlov:1999qy}.  However, such jets are generated only
in fast rotators and may not be generic
\cite{Ott:2006wd,Burrows:2007qe,Dessart:2007kl}.  Another possibility
is anisotropic neutrino emission from the cooling proto-neutron star.
If strong magnetic fields are present, neutrino-matter interactions
can generate dipole asymmetries of $\sim$$1$\%, leading to recoil on
the order of a few hundred km$\,$s$^{-1}$
\cite{Lai:1998tg,Nardi:2001hc,Lai:2001ij,Socrates:2005fr}.  These
scenarios require magnetar field strengths (i.e. $10^{14}-10^{15}$ G)
and/or exotic neutrino physics
\cite{Kusenko:1999kx,Lambiase:2005yq,Barkovich:2004vn,Fuller:2003yq}
and may not produce substantial kicks in typical core-collapse
supernovae.

If neutron star kicks are a generic feature of core collapse, then the
most natural explanation is recoil due to an asymmetric supernova
explosion
\cite{Burrows:1996th,Scheck:2004rt,Burrows:2007eu,Scheck:2006vn}.
During axisymmetric core collapse, the stalled bounce shock is unstable to neutrino-driven 
convection and low-order $\ell$-modes.  Significant
asymmetry at the onset of neutrino-driven shock revival should
naturally lead to an asymmetric explosion and the hydrodynamic recoil
of the PNS
\cite{Janka:1994fk,Burrows:1996th,Scheck:2004rt,Fryer:2004kx,Scheck:2006vn}.

Observations of large-scale asymmetries in young supernova remnants lend qualitative support to the hydrodynamic mechanism \cite{Kjaer:2010lr}. Unfortunately, multi-dimensional,
radiation-hydrodynamic simulations of recoil are computationally
challenging.  A proper study requires simulating the full physics of
collapse, the formation of the PNS, the development of
instabilities during the post-bounce phase, the evolution of the
asymmetric explosion, the off-axis movement of PNS, and the full
decoupling of the ejecta from the PNS.  Because the expanding
post-shock material interacts with the PNS through both pressure
and gravity, this requires following the shock out to 
large distances (hundreds of thousands of kilometers) and
late times (several seconds).  Complicating matters is that during
this evolution, one must continue to resolve the movement of the
PNS and the surrounding highly nonlinear flow.  

\citeauthor{Scheck:2006vn}~2006 present a practical approach to this problem
\cite{Scheck:2004rt,Scheck:2006vn}.  By excising the PNS and
replacing it with a rigid, contracting boundary, they avoid severe
Courant timestep restrictions.  They also greatly simplify their
radiation transport, enforcing a constant luminosity at their inner
boundary, and begin their calculations 20 ms after bounce.  These
approximations allow \citeauthor{Scheck:2006vn}~to follow the
evolution of the shock to large distances and late times, and to
perform a detailed parameter study.  Unfortunately, this approach
requires them to infer a kick through a rigid, impenetrable boundary.
Their results should therefore be checked by more realistic (though
costly) simulations.  

As a complement to the work of \citeauthor{Scheck:2006vn}, we present
a two-dimensional (2D) simulation of the collapse of a 15-$M_\odot$
progenitor core.  By employing a pseudo-Cartesian mesh at the center
of our domain, we naturally capture the neutron star's formation and
any subsequent off-center acceleration.  During our simulation, the
proto-neutron star forms, after which it recoils due to a delayed, neutrino-driven, anisotropic explosion.
The explosion is artificially induced by adding additional neutrino luminosity to the calculation.  At the
end of our simulation, the NS has achieved a velocity of $\sim$150 km
s$^{-1}$ and is still accelerating at $\sim$350 km$\,$s$^{-2}$.  The
recoil is primarily hydrodynamic in nature, with anisotropic neutrino
emission contributing less than 2\% of the overall kick magnitude.
Most notably, we obtain a significant kick without invoking strong
magnetic fields, exotic neutrino physics, or misaligned jets.  Our
results are consistent with the previous \citeauthor{Scheck:2006vn}~studies \cite{Scheck:2004rt,Scheck:2006vn}.
Taken together, these simulations provide compelling
numerical support for the hydrodynamic mechanism of neutron star
kicks.

\begin{figure}[ht!]
\begin{center}
\includegraphics[width=\fullfigwidth,angle=0,clip=true]{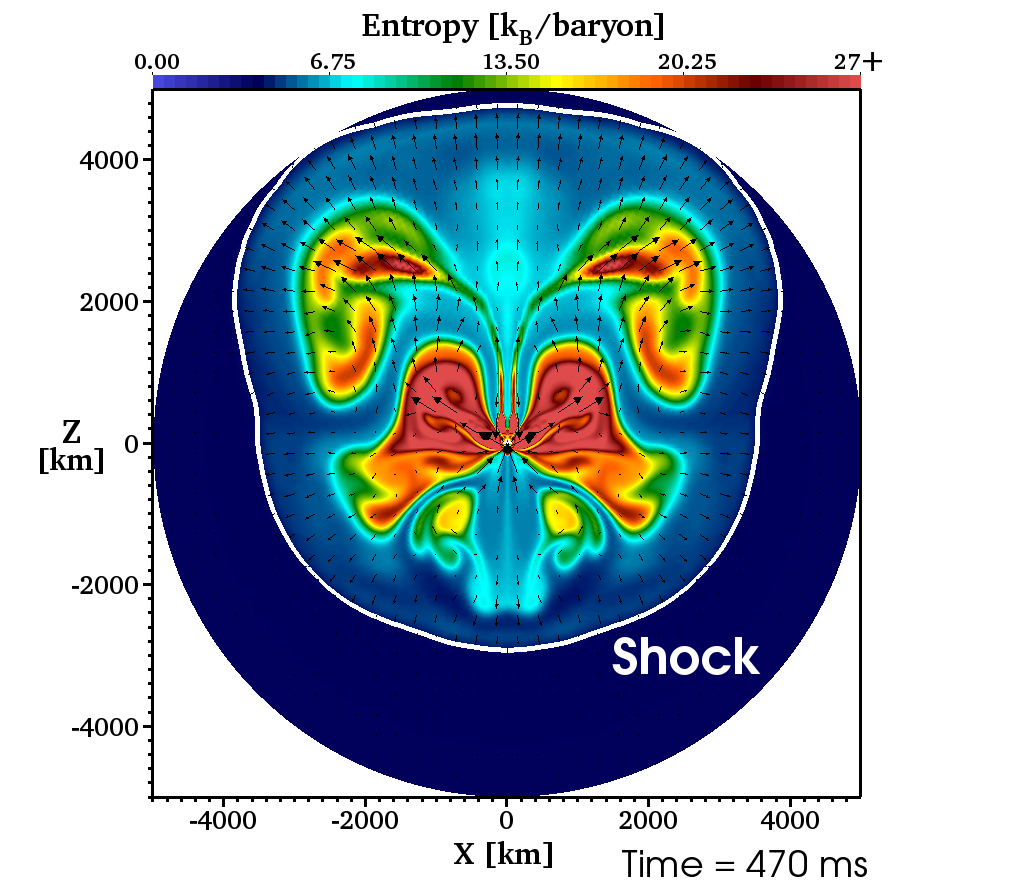}
\includegraphics[width=\fullfigwidth,angle=0,clip=true]{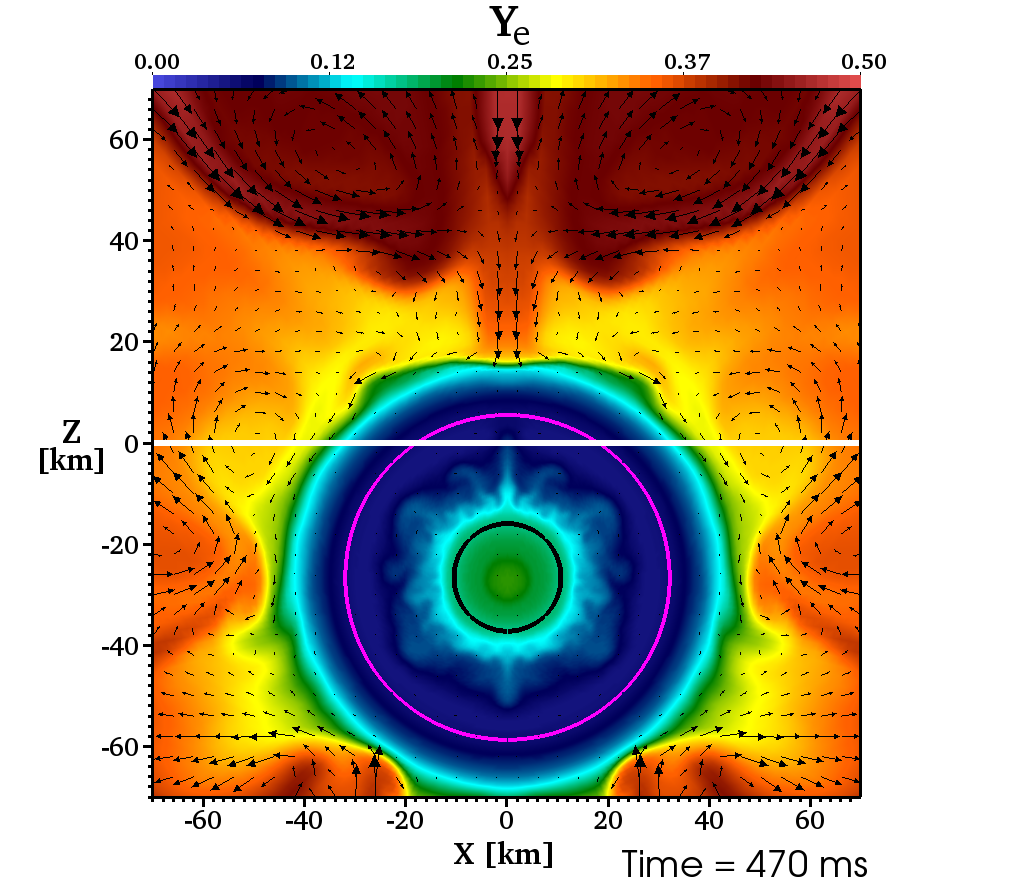}
\caption{The recoil of the proto-neutron star due to an asymmetric
  core-collapse supernova explosion.  The large-scale explosion is
  primarily in the $+Z$ direction (top) while the PNS is kicked
  in the $-Z$ direction (bottom).  In the bottom panel, the white line
  denotes $Z=0$, while the purple and black curves represent the isodensity
  surfaces where $\rho=10^{12}$ and $10^{14}$ g cm$^{-3}$,
  respectively.  Velocity vectors are overlaid in black.}
\label{fig:explosion}
\end{center}
\end{figure}

\section{\label{sec:setup} Numerical Setup and Methods\protect}
Our 2D, axisymmetric
calculations are performed with the multi-group, arbitrary
Lagrangian-Eulerian (ALE), radiation-hydrodynamics code {\sc VULCAN/2D}
\cite{Livne:1993zr}.  We perform 2D radiation transport using the
multi-group flux-limited diffusion approximation
\cite{Livne:2004zr}.  We simulate the collapse of the inner 5000 km of a
non-rotating, 15-$M_\odot$, solar-metallicity, red-supergiant
progenitor \cite{Woosley:1995mz}.  Exterior to 20 km, our
computational domain is a spherical-polar mesh which transitions to a
pseudo-Cartesian grid in the center.  Such a grid avoids severe
timestep restrictions due to the convergence of angular zones and
frees the PNS to move in response to radiation or hydrodynamic
forces.  Our mesh covers the full 180$^\circ$, 2D domain with 120
angular zones and 330 radial zones (logarithmically spaced exterior to
the inner Cartesian region).  We employ the finite-temperature nuclear
equation of state of \citeauthor{Shen:1998gf} \cite{Shen:1998gf,Shen:1998ve} and include
self-gravity with a grid-based solution of the Poisson equation
\cite{Burrows:2007ly}.  To ensure that we optimally resolve the
high-density core, we allow our grid to track the PNS.  Our
remapping scheme determines the center of mass of the inner core
(i.e.~densities above $10^{12}$ g cm$^{-3}$) after each timestep and
shifts the mesh to keep the core centered while ensuring momentum conservation.

Despite decades of intense theoretical effort, the success of the
delayed-neutrino mechanism
\cite{Colgate:1966rt,Wilson:1985ys,Bethe:1985fr} in driving
core-collapse supernova explosions has still not been demonstrated
\cite{Burrows:1995zr,Mezzacappa:2001mz,Fryer:2004ly,Janka:1996vn,Janka:2007gf,Blondin:2003ul,Buras:2006qf,Burrows:1993pd,Liebendorfer:2005lq,Marek:2009lr,Suwa:2009lr}.
However, recent calculations have shown that this mechanism's
capacity to power explosions increases with dimension
\cite{Murphy:2008dq,Nordhaus:2010fr}.  Ambitious three-dimensional
calculations with accurate neutrino transport may yet validate the
delayed-neutrino mechanism.

Because previous core-collapse studies with {\sc VULCAN/2D} did not
produce neutrino-driven supernovae
\cite{Ott:2008cr,Ott:2006wd,Dessart:2006nx,Burrows:2007eu,Brandt:2010xy}, we induce
explosions by supplementing the radiation transport with additional
electron and anti-electron neutrino luminosity ($L_{\nu_e} =
L_{\overline{\nu}_e}= 2\times10^{52}$ erg$\,$s$^{-1}$) as described in
\cite{Murphy:2008dq,Nordhaus:2010fr}.  This represents an enhancement
in the $\nu_e$ and $\overline{\nu}_e$ luminosities of
$\sim$$50$\%.  The core collapses to
nuclear densities, launching a bounce shock which stalls and is
subsequently revived mainly by charged-current neutrino absorption after a delay of approximately 135 milliseconds.

\section{\label{sec:recoil}Recoil from Asymmetric Core-Collapse Explosions\protect}
At the onset of explosion, the hydrodynamic flow behind the shock is
turbulent and the shock itself is deformed by the development of
low-mode instabilities
\cite{Blondin:2003ul,Scheck:2004rt,Scheck:2006vn,Blondin:2007mz,Fernandez:2010ly}.
The PNS recoils due to the blast's anisotropic
propagation through the stellar envelope.  We follow the explosion and the
acceleration of the PNS until 470 ms after bounce, at which point
the shock front reaches the boundary of our computational domain (5000 km).
Figure $\ref{fig:explosion}$ shows the global explosion geometry and
the position of the PNS at the end of our calculation.  The top
panel is an entropy map of our computational domain with velocity
vectors overlaid and the shock position outlined in white.  The bottom
panel shows the electron fraction $Y_{\rm e}$ over the inner $\sim$70
km.  The white line is the $Z=0$ axis, while the pink and black curves
represent the $10^{12}$ g cm$^{-3}$ and $10^{14}$ g cm$^{-3}$
isodensity contours, respectively.  Note that the asymmetry of the
explosion in the $+Z$-direction leads to a PNS recoil in the
$-Z$-direction.  While axisymmetry restricts our core to motion along the $Z$-axis,
three-dimensional computations would impose no such constraint and
could produce a recoil in any direction for initially non-rotating
progenitors.  Note that the presence of rotation may lead to a
preferred explosion direction and, hence, kick direction.  The
differences between kicks from non-rotating and rotating progenitor
models should be investigated in 3D.

While VULCAN/2D automatically and self-consistently computes the acceleration of the
core, it does not output the individual forces governing the motion of the
PNS.  We therefore post-process our results by computing the
hydrodynamic acceleration $\vec{a}_{\rm c}$ of the core due to
anisotropic gravitational forces, pressure forces, and momentum flux.
The Eulerian equations of hydrodynamics give
\begin{equation}
\vec{a}_{\rm c}=\dot{\vec{v}}_{\rm c}\sim\int_{r> r_{\rm
    c}}\frac{G\vec{r}}{r^3}dm - \frac{1}{M_{\rm
    c}}\left[\oint_{r=r_{\rm c}} P d\vec{S} + \oint_{r=r_{\rm c}}\rho v_{\rm r}\vec{v}dS \right],
\label{eq:acceleration}
\end{equation}
where $\rho$ is the density, $M_{\rm c}$ and $\vec{v}_{\rm c}$ are the mass and mean velocity
of the inner region (where $\rho\geq10^{12}$ g$\,$cm$^{-1}$), $P$ is the gas pressure,
$\vec{v}$ is the fluid velocity, $v_{\rm r}$ is the radial component
of the velocity, and $r_\mathrm{c}$ is a fiducial spherical radius.  The code self-consistently yields the recoil speed of the PNS (approximately bounded by the purple curve in Fig.~\ref{fig:explosion}), but we can use Eq.~\ref{eq:acceleration} to determine the various contributions to its acceleration and consequent motion.

\begin{figure}[ht!]
\begin{center}
\includegraphics[width=\smallfigwidth,angle=0,clip=true]{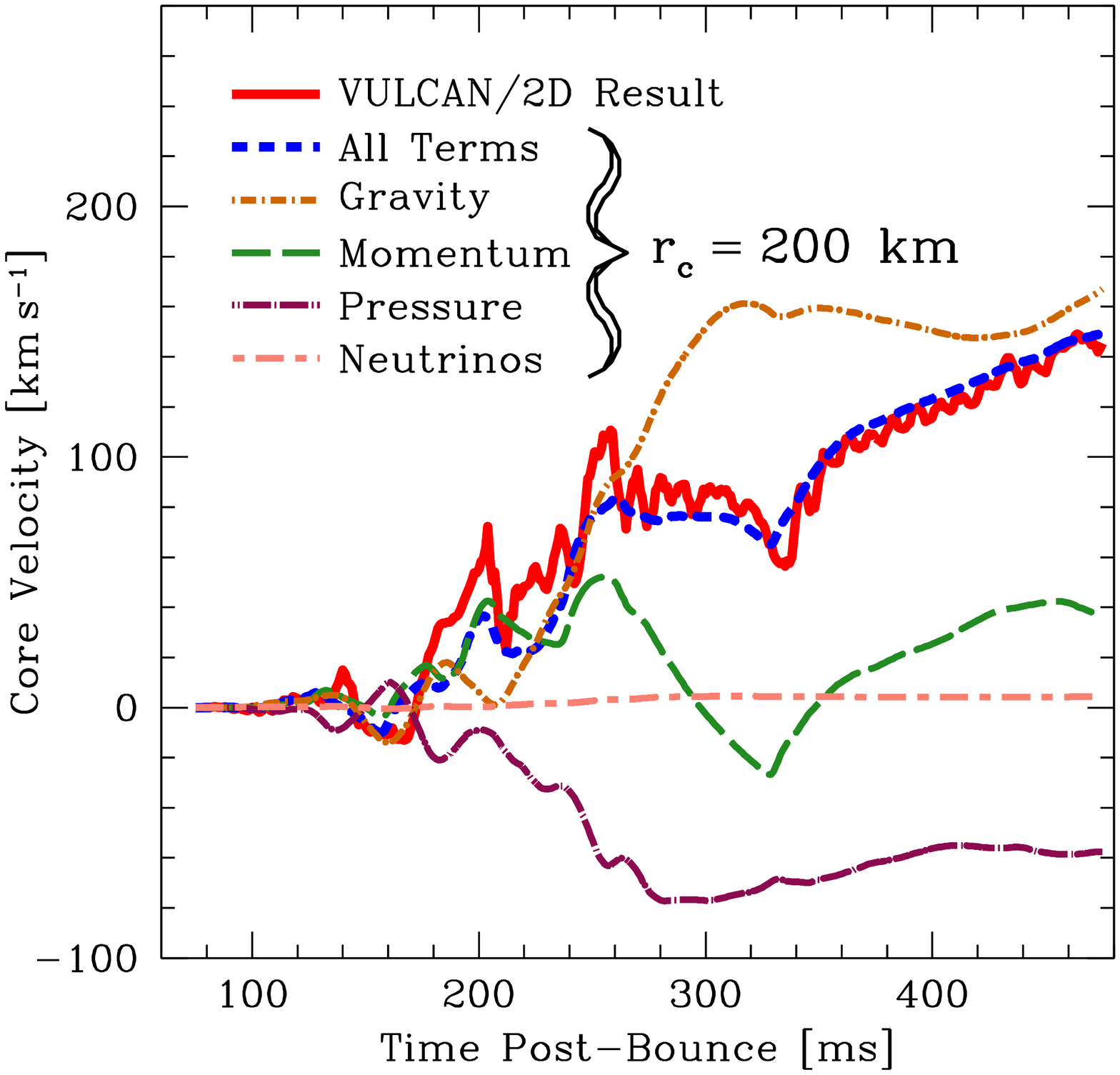}
\includegraphics[width=\smallfigwidth,angle=0,clip=true]{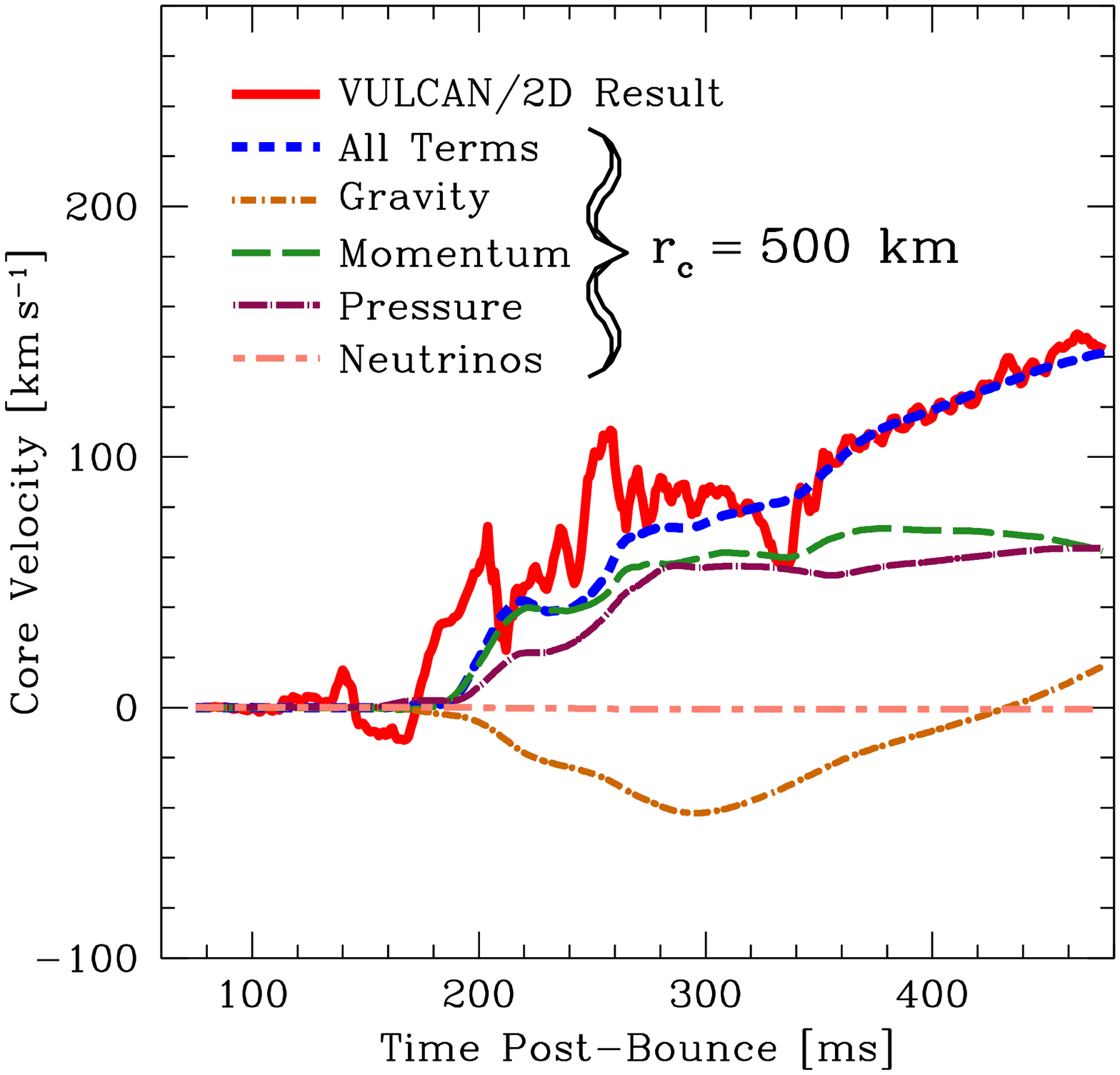}
\caption{The core velocity as a function of time after bounce.  The
  solid-red curve in both figures shows the core velocity, in the $-Z$ direction, as a
  function of time after bounce in our simulation.  Though the
  inferred core velocities calculated at $r_{\rm c}=200$ km (top figure) and
  $r_{\rm  c}=500$ km (bottom figure) accurately reproduce the actual core
  velocity at late times, this figure demonstrates that one must
  exercise caution when interpreting the relative contribution of each
  component.  Anisotropic neutrino flux contributes very little
  ($\lesssim 2$\%) of the total kick at all radii.}
\label{fig:kick}
\end{center}
\end{figure}

The first term in Eq.~\ref{eq:acceleration} represents the
acceleration due to the gravitational field exterior to $r_{\rm c}$,
assuming a spherically symmetric distribution of matter interior to
this radius.  The second term is due to anisotropic gas pressure,
while the third term represents the contribution due to momentum flux.
In a spherically symmetric explosion, each term would vanish
individually.  These three terms include all hydrodynamic forces, but
do not include asymmetries in the radiation pressure.  In our simulations, exterior to the
radius at which the flux limiter transitions to free-streaming,
anisotropic neutrino momentum contributes $\sim$2\% of the total kick
(see Fig. \ref{fig:kick}).  

In general, the relative contributions of the various terms in
Eq.~\ref{eq:acceleration} will depend sensitively on the
radiation-hydrodynamics and explosion dynamics.  For instance, a
spherically-symmetric distribution of ejected mass possessing
asymmetric ejection velocities will lead to gravity and momentum terms
of the same sign.  In particular, since the PNS recoils towards the lower-velocity ejecta, the gravitational acceleration is in the same direction as the kick.  This gravitational ``tug-boat" effect enhances the recoil.  Isotropic ejection velocities with anisotropic mass
loss results in the gravity component partially canceling the momentum
contribution.

We present the PNS kick velocity (as computed by VULCAN/2D) as
a solid red line in both the top and bottom panels of
Fig.~\ref{fig:kick}.  Using Eq.~\ref{eq:acceleration}, we show the inferred kick velocity (dashed-blue
curve) and its components at
200 km (top panel) and 500 km (bottom panel).  These curves
represent the mean velocities of matter interior to 200 km and 500 km.  
As the core evolves, matter
interior to 500 km becomes more centrally concentrated and its average
velocity approaches that of the innermost regions (i.e.~the monopole approximation gets better and better).  The agreement
between the red line and the blue lines therefore improves with time.

Figure \ref{fig:kick} demonstrates that the kick imparted to the
PNS may be inferred by evaluating Eq.~\ref{eq:acceleration}
even at large radii.  However, the relative contributions of the three
terms in Eq.~\ref{eq:acceleration} differ dramatically.  At $r_{\rm
c}=200$ km, the late time evolution of our simulation is dominated by the gravitational
component, while the momentum and pressure contributions are of
opposite sign and comparable in magnitude.  For $r_{\rm c}=500$ km,
the pressure and momentum contributions are approximately equal (in
both sign and magnitude) and nearly constant between $\sim$200 ms and
$\sim$470 ms.  The secular evolution of the PNS velocity at the
end of our calculation is governed by the gravitational component.
The one component which does not depend strongly on radius is the
contribution from anisotropic neutrino emission, which is small
($\lesssim$2\% of the kick).

The interpretation of the kick (though not its value) thus
depends on the radius at which the terms of Eq.~\ref{eq:acceleration}
are evaluated.  At large radii, pressure and gravity vanish and an
observer will attribute the entire kick to anisotropic momentum flux.
The story is very different near the PNS itself.  Because the
inner core is nearly in hydrostatic equilibrium, pressure and
gravity are both very large and in balance.  An observer in
this region would remark on the near cancellation of the
gravitational and pressure terms in Eq.~\ref{eq:acceleration}.  For
example, in our calculations, with a radius $r_\mathrm{c}$ that moves
inward to always enclose 1.3 $M_\odot$, these two components of the
kick cancel to one part in $10^2$.  Our results demonstrate the
limitations of interpreting the individual components of
Eq.~\ref{eq:acceleration}.  Since pressure and gravity do work on
expanding matter, their contributions to the acceleration decrease in
magnitude relative to the contribution due to the anisotropic momentum
flux.  

\subsection{\label{subsec:extrapolation}Extrapolating the Kick\protect}
\begin{figure}[ht!]
\begin{center}
\includegraphics[width=\fullfigwidth,angle=0,clip=true]{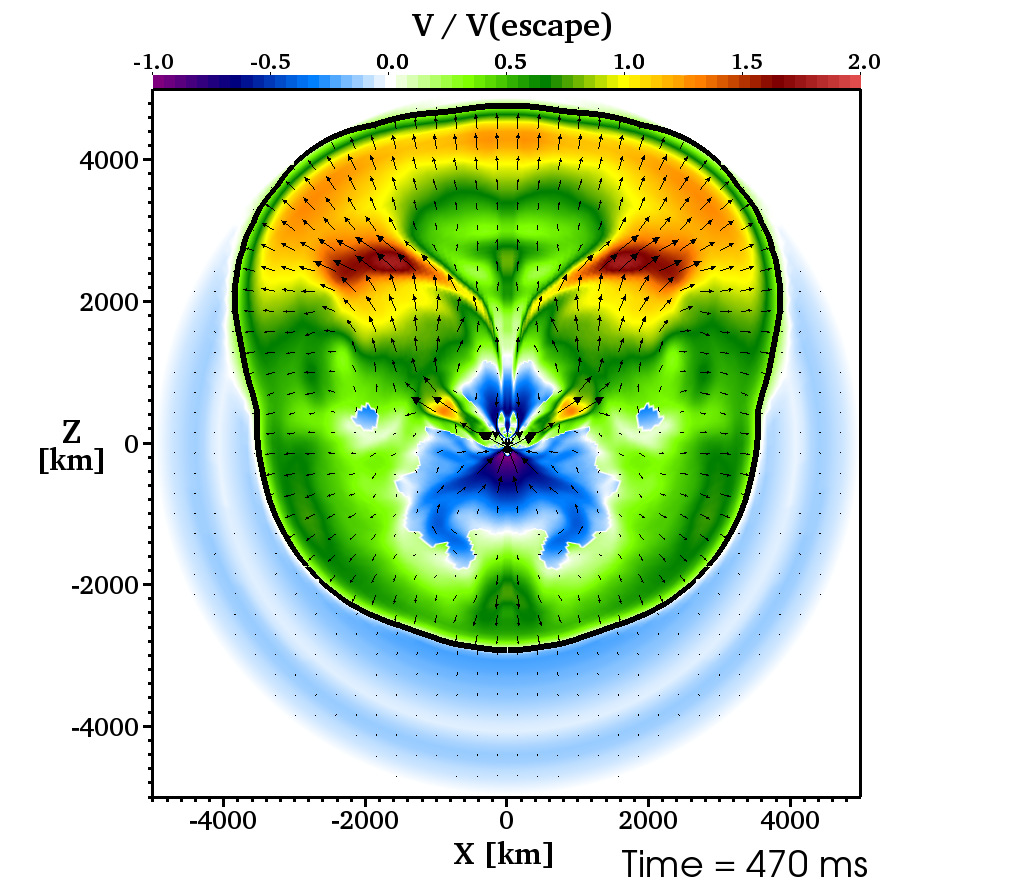}
\includegraphics[width=\fullfigwidth,angle=0,clip=true]{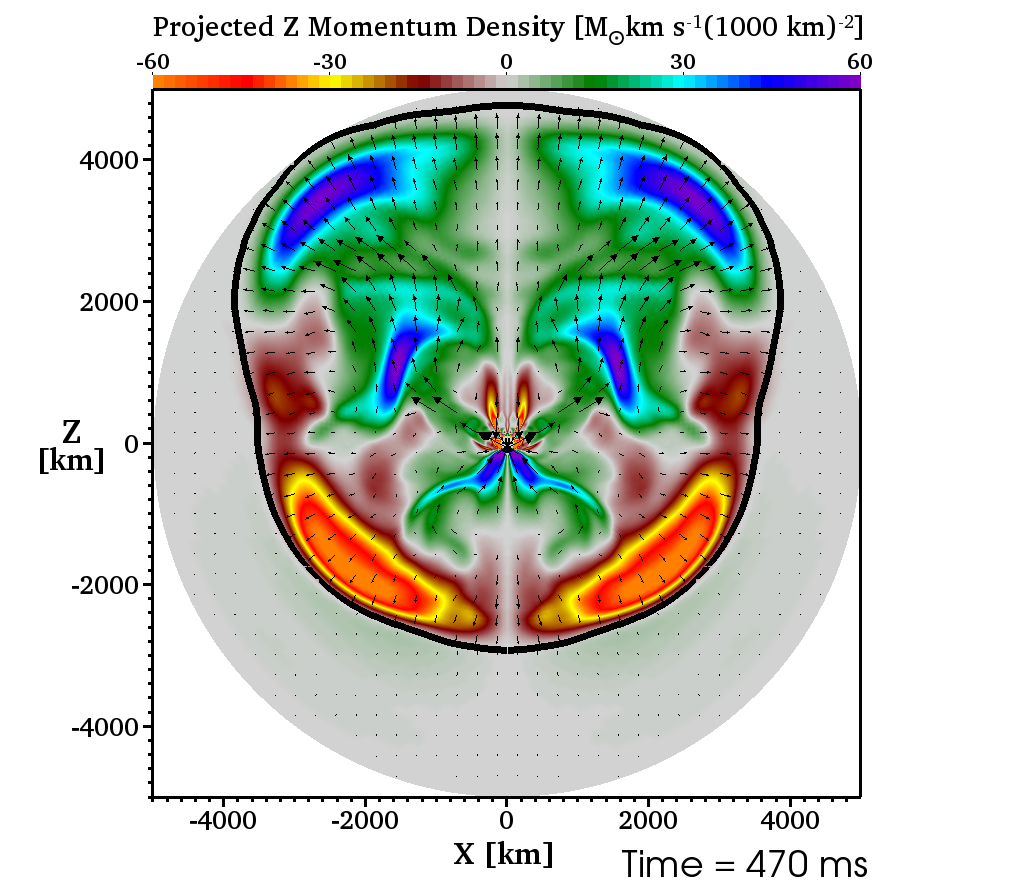}
\caption{Top: The ratio of the fluid velocity, $v$, to escape speed,
  $v_{\rm esc}$, as a function of position 470 ms after bounce.
  Bottom: Projected $Z$-momentum density $p_Z$ as a function of time
  470 ms after bounce.  The cylindrical volume element is included, so
  that $\int p_Z \ud X \ud Z$ gives the total $Z$-momentum.  We have
  overlaid velocity vectors and a thick black curve representing the
  position of the shock on both panels.}
\label{fig:vesc}
\end{center}
\end{figure}

Figure~\ref{fig:kick} indicates that our PNS is still
accelerating at $\sim$350 km$\,$s$^{-2}$ when the shock has reached
the boundary of our computational domain.  However, the ejecta have
not yet decoupled from the core to reach the ballistic regime.  The
spatial distributions of momentum and velocity offer a hint of the
core's future evolution, but unfortunately do not permit a
straightforward extrapolation.  Ideally (though at considerable
computational expense), this would be handled by remapping our results
onto a larger grid and continuing a full radiation-hydrodynamic calculation.
However, momentum and velocity maps, which we show in
Fig.~\ref{fig:vesc}, offer a useful picture of the ejecta at the end
of our calculation.

The top panel of Fig.~\ref{fig:vesc} shows the velocity of matter
throughout our computational domain in units of the local escape
speed, calculated assuming a spherically symmetric distribution of
matter.  Because the potential is dominated by the PNS, this approximation is extremely accurate.  The map
clearly shows that our model has not yet reached the ballistic regime,
and that the matter behind the shock is still accelerating and
evolving dynamically.  A significant region of matter at $Z \sim
-1000$ km seems likely to fall back, while a pocket of material at $Z
\sim 2500$ km is expanding at nearly twice the local escape speed.
The infalling region has only $\sim$20\% of the momentum in the core
and, thus, is unlikely to significantly affect our inferred kick.
However, the complexity of the hydrodynamics makes it impossible to
extrapolate by assuming, for example, self-similar expansion.

The lower panel of Fig.~\ref{fig:vesc} shows the projected
$Z$-momentum density, $p_Z \equiv \pi R \rho v_Z$.  The factor $\pi
R$, where $R$ is the cylindrical radius, is the length of a semicircle
of revolution.  This projects the half-cylinder defined by
$0<\phi<\pi$ in 3D onto the half-plane $X>0$ in 2D, so that $\int p_Z
\ud X \ud Z$ gives the correct value for the total $Z$-momentum.  This
map shows that the high-velocity bubbles at $Z \sim 2500$ km are
regions of low density; most of the momentum is concentrated behind
the shock and in the regions behind the highest velocity ejecta at $Z
\sim 1000$ km.  At the end of our calculation the PNS is still
injecting mass and momentum into these regions.  There appears to be no such
injection of momentum into the regions at negative $Z$.  If this
causes the expansion of matter to slow in the $-Z$ direction, it could
help maintain an asymmetric matter distribution, and thus the
gravitational component of its acceleration, for several seconds.

The continued acceleration of the PNS will depend on the
evolution of the asymmetry of shocked material.  There are a variety
of ways to quantify this asymmetry, as discussed in
\cite{Scheck:2006vn,Burrows:2007ly}.  We choose $\alpha \equiv \langle
v_z \rangle / \langle |v| \rangle$, where $\langle \rangle$ denotes a
mass-weighted average over the post-shock region with $r > 100$ km (to
exclude the PNS itself).  This is similar to the $\alpha$
presented in \cite{Scheck:2006vn}.  If we assume this asymmetry to be constant in time, material on one side of the PNS will be a factor of $1-\alpha$ as close as material on the other side.  We may then crudely estimate
the gravitational acceleration of the core, $a_{c,{\rm grav}}$ as
\begin{equation}
a_{c,{\rm grav}} \sim
G M_{\rm sh} \left( \frac{1}{\left[(1-\alpha)r_{\rm sh}\right]^2} -
      \frac{1}{r_{\rm sh}^2} \right)
 \approx \frac{2 \alpha G M_{\rm sh}}{r_{\rm sh}^2}
\end{equation}
for small $\alpha$, where $r_{\rm sh}$ is the shock radius and $M_{\rm sh}$ is the total mass of ejecta
and shocked envelope material.  In our calculation, $\alpha \sim 0.1$
from 300 milliseconds to 470 milliseconds after bounce.  Assuming $M_{\rm sh} \sim M_\odot$ and $\alpha \sim
0.1$, then for $a_{c,{\rm grav}}$ to be of order 1 km$\,$s$^{-2}$, we
need to follow the shock out to $\sim$10$^5$ km.  This corresponds to
5 seconds at a shock velocity of 20,000 km$\,$s$^{-1}$, and represents
a challenging computational problem.  We hope ultimately to address
this problem with CASTRO \cite{Almgren:2010fk,Nordhaus:2010fr}, a new adaptive mesh
refinement radiation-hydrodynamics code, which will allow us to follow the shock
while still resolving the PNS.  

\subsection{\label{subsec:comparison}Comparison to Previous Work\protect}
Our approach of following the collapse of a massive star's core, the
formation of a natal PNS, and the subsequent off-axis motion
complements previous studies that infer kicks on an excised PNS
\cite{Scheck:2004rt,Scheck:2006vn}.  By omitting the inner regions,
starting the simulation $\sim$20 ms after bounce, and imposing a
constant inner neutrino luminosity, \citeauthor{Scheck:2006vn}~greatly
reduced the problem's computational cost.  They were thus able to
follow the shock evolution to large distances ($>10^4$ km) and late
times ($>$1 s).  To approximate a physical neutron star, those authors
used a contracting inner boundary motivated by radiation-hydrodynamic
simulations \cite{Scheck:2006vn}.  While attractive for calculating
long-term evolution, their approach requires one to infer a PNS
kick through a rigid boundary of infinite inertial mass.  This
assumption neglects effects resulting from displacement of the
PNS relative to the surrounding fluid.  To compensate, in a
subset of their simulations, these authors artificially add the
inferred kick velocity to the gas, mimicking movement of the PNS.
Our work handles all of these effects self-consistently, providing an
important check on the various approximations made in
\cite{Scheck:2004rt,Scheck:2006vn}.

Another difference between our work and that of
\citeauthor{Scheck:2006vn}~is that we implement the momentum equation
in conservative form using a grid-based solution to the Poisson
equation.  As a result, our model conserves total momentum to better
than 1\% of the core's final value.  \citeauthor{Scheck:2006vn}~solve the
Poisson equation using a Legendre expansion with a relativistic
correction \cite{Rampp:2002lr,Marek:2006fk,Scheck:2006vn}.  Recently, \citeauthor{Wongwathanarat:2010lr}~performed a three-dimensional study using the same techniques in the \citeauthor{Scheck:2006vn} two-dimensional studies and arrived at similar conclusions.  

Given the differences in our complementary techniques, the agreement
between our results and those of \citeauthor{Scheck:2006vn} is
gratifying.  Our detailed calculations of the first few hundred
milliseconds including the core support the work of
\cite{Scheck:2004rt,Scheck:2006vn}, while their extended calculations
indicate that a final kick magnitude of at least 400-500
km$\,$s$^{-1}$ may be likely for our model.  Taken together, this body
of work strongly supports the case that asymmetric supernova
explosions lead naturally to substantial recoil of the PNS.

\section{\label{sec:conclusions}Conclusions\protect}
In this work, we have presented the first multi-dimensional,
multi-neutrino-energy-group, radiation-hydrodynamic simulation of a core-collapse
supernova that results in a formation and acceleration of a nascent
neutron star.  The recoil of the PNS naturally arises from the
asymmetric nature of the neutrino-driven explosion.  At the end of our
simulation the PNS has reached a velocity of $\sim$150 km
s$^{-1}$, but is still accelerating at $\sim$350 km$\,$s$^{-2}$.  While it
is difficult to extrapolate the acceleration to later times, our
PNS would need to maintain this value for only a few hundred
milliseconds more to reach the peak of the observed pulsar velocity
distribution.  This is suggested by Fig.~\ref{fig:vesc}; the continued
ejection of momentum in the $+Z$-direction could maintain the
asymmetric matter distribution and continue to gravitationally
accelerate our PNS.  It should also be noted that the highest observed kicks (those upwards of 1000 km s$^{-1}$) may result from the most asymmetric and energetic explosions.

Hydrodynamic recoil due to neutrino-driven, core-collapse supernovae provides a
natural mechanism for accelerating neutron stars and pulsars without
the need to appeal to anisotropic neutrino emission or more exotic scenarios.  However, a definitive
confirmation of this mechanism will require a self-consistent model of
core-collapse supernova explosions.  To avoid constraints imposed by
axisymmetry, future work should investigate recoil and explosion
anisotropies in three dimensions and compare the resulting kick
velocities with observations.

\begin{acknowledgments}
J.N. and A.B. are supported by the Scientific Discovery through
Advanced Computing (SciDAC) program of the DOE, under grant number
DE-FG02-08ER41544, the NSF under the subaward ND201387 to the Joint Institute for Nuclear Astrophysics (JINA, NSF PHY-0822648), and the NSF PetaApps program, under award OCI-0905046 via a subaward 44592 from Louisiana State University to Princeton University.  Computational resources were
provided by the TIGRESS high performance computer center at Princeton
University, the National Energy Research Scientific Computing Center
(NERSC; under contract DE-AC03-76SF00098), and on the Kraken and
Ranger supercomputers, hosted at NICS and TACC via TeraGrid award
TG-AST100001.  This material is based upon work supported under a
National Science Foundation Graduate Research Fellowship to T.D.B. C.D.O. is partially supported by the NSF under grant numbers AST-0855535 and OCI-0905046.
\end{acknowledgments}

\bibliography{nordhaus}

\end{document}